# Towards the Internet of X-things: New Possibilities for Underwater, Underground, and Outer Space Exploration

**Written By: Nasir Saeed, Mohamed-Slim Alouini, and Tareq Y. Al-Naffouri, King Abdullah University of Science and Technology, Thuwal, Makkah, Saudi Arabia.**

The rapid growth of the world's population demands more natural resources, food, and space. World renowned physicist Stephan Hawking has argued that soon we will require another world to live on because we are running out of space and natural resources. This ever-increasing demand for resources and space needs novel technologies to explore natural resources, produce more crops, and explore outer space. Internet of X-things (X-IoT) is an enabling technology to overcome all of the above issues. The framework of X-IoT consists of three major categories as shown in Figure 1. The first one is the Internet of underwater things (IoUT) for smart oceans. The second category is the Internet of underground things (IoUGT) for smart agriculture, seismic monitoring, and Oil/Gas fields. The third category is the Internet of space things (IoST) for outer space exploration, to provide global coverage, and to enable inter-satellite communications. X-IoT is made up of billions of smart objects from Nano-sensors to large machines communicating in underwater, underground, and outer space as illustrated in Figure 1. The breathtaking growing rate of using these smart objects will lead to the development of smart oceans, smart oil fields, smart cities, smart agriculture, and smart outer space soon. The importance of X-IoT can be seen from the investment in this technology where its total worth is expected to reach 6.2 Trillion US Dollars by 2025 [1]. Recently, the research on X-IoT has shown that it provides various underwater, underground, and space applications. For example, X-IoT can provide smart marine exploration and monitoring solutions, provide better food production and supply-chain management solutions in agriculture, and improve the production of oil and gas fields.

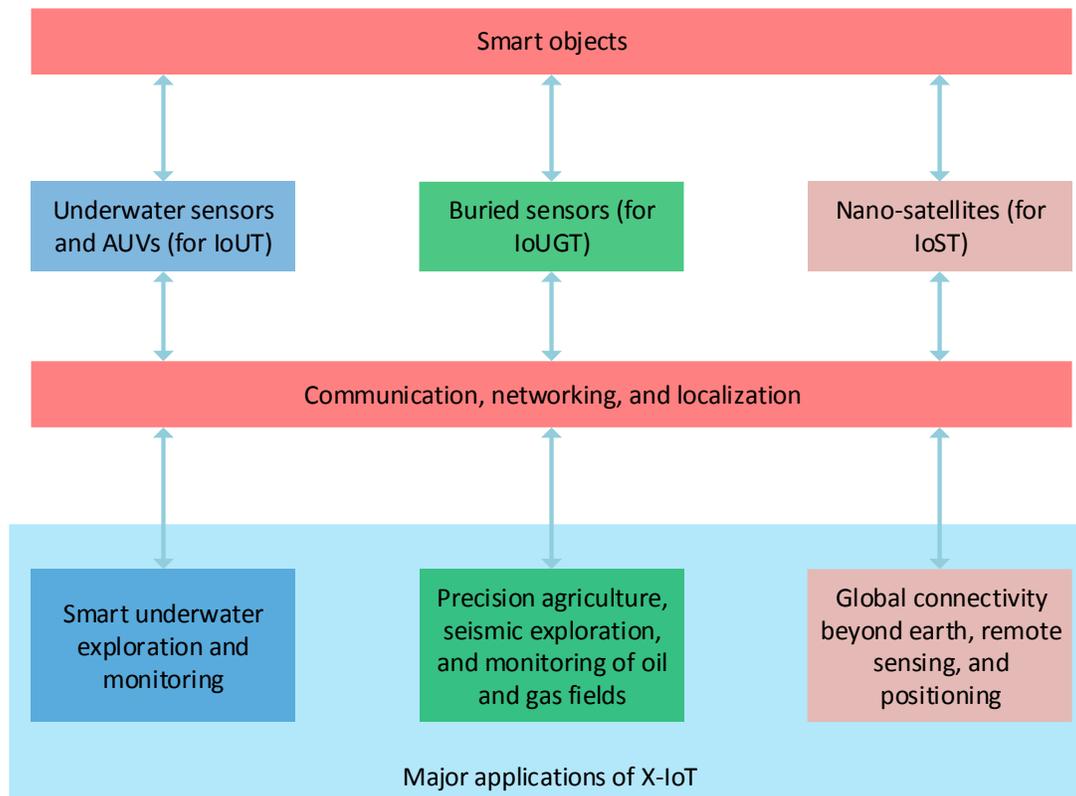

**Figure 1 Illustration of X-IoT framework.**

## Internet of Underwater Things (IoUT):

Oceans, which occupy 71% of planet Earth surface, produce numerous benefits to humankind, e.g., food supply, climate regulation, recreation, medicine, and transportation. Hence, oceans based businesses contribute more than 500 billion US dollars to the world economy [2]. Nevertheless, 95% of oceans remains unexplored, which necessitates deploying underwater smart objects such as sensors, remotely operated vehicles, and autonomous underwater vehicles, as a means to explore and monitor the oceans. The goal of IoUT is to create a worldwide network of smart underwater objects which connect the oceans, lakes, and streams to enable various applications such as climate regulation, the study of marine animals, the monitoring of underwater oil rigs, intruder detection, and unmanned operations. Figure 2 shows a possible architecture for IoUT based on underwater optical wireless communications which consists of smart objects such as underwater sensors and autonomous underwater vehicles (AUVs). Optical base stations (OBS) are used to connect the smart underwater objects to surface station by using optical links [3].

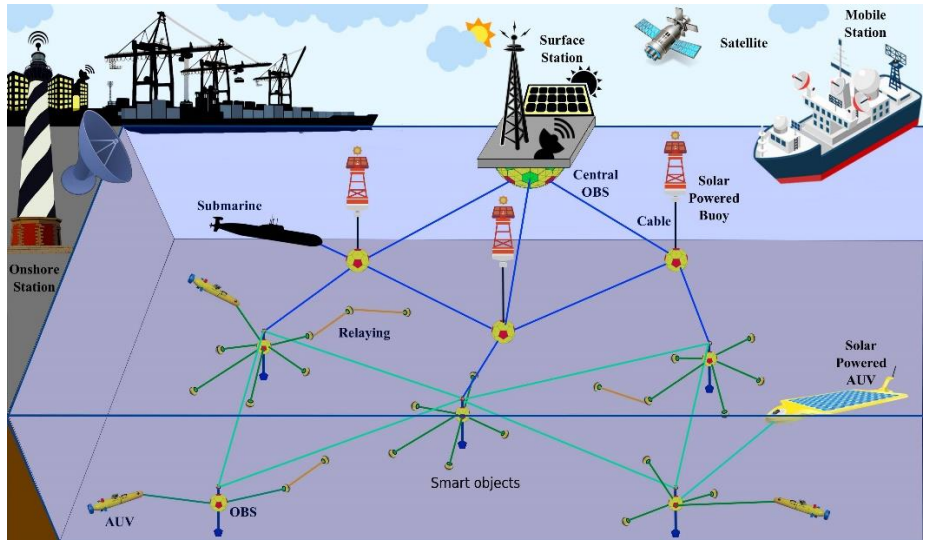

Figure 2 Architecture of optical wireless communication-based IoUT.

## Technological Challenges:

The technological challenges for IoUT can be broadly classified into communication, networking, and localization.

### 1) Communication Challenges:

IoUT applications require a medium to communicate in the underwater environment and from this environment to the outside world. Hence, communication in the IoUT is of paramount importance, however, it poses a series of challenges. For example, radio waves suffer from high absorption due to their high frequencies and can only travel for few meters on the surface of the water. Alternatively, acoustic waves provide long transmission range, but due to the low speed of sound waves, the data rate is very low. Therefore, optical waves have recently been introduced to provide high data rate for underwater communications. However, optical waves in underwater suffer from low transmission range and require accurate pointing and acquisition mechanisms. Table 1 shows the advantages and disadvantages of potential communication technologies for IoUT [3].

| Features | Acoustic waves | RF waves | Optical waves |
|---|---|---|---|
| Data rate | In Kbps | In Mbps | In Gbps |
| Bandwidth | 100 Hz – 1 KHz | In MHz | 10-150 MHz |
| Transmission power | In tens of Watts | Up to 100 of watts | Up to 1 Watt |
| Transmission distance | In tens of Km | In tens of meters | 100-200 m |
| Latency | High | Moderate | Low |

Table 1 Features of various wireless communication technologies for IoUT.

2) Networking and Localization Challenges:

Besides the type of medium for communication, networking is also a challenging task for IoUT due to the harsh and dynamic underwater environment. Although research has been carried out to develop networking protocols for acoustic-based IoUT, few works exist for networking in optical communication-based IoUT such as [4] and the references therein. Furthermore, due to the non-availability of GPS signals under the water, localization of the smart objects in IoUT is an important requirements implementation of IoUT. For instance, in one of the recent works, we have estimated the location of sensors with limited connectivity for optical communication based underwater sensor networks by using multidimensional scaling [5]. Another example of localization for optical communication-based IoUT includes the study of outliers and anchors geometry on localization accuracy. Underwater optical channel impairments such as absorption, scattering, turbulence, and air bubbles can cause large range measurement errors (outliers). Therefore, in [6], we have tackled the problem of outliers to improve the overall accuracy of the localization. Also, every location finding method depends on the geometry of the anchors, and if the geometry of the anchors is not optimized, it can lead to a large localization error. Therefore, we have optimized the anchor's geometry to better estimate the location of underwater sensors.

Internet of Underground Things (IoUGT):

The ever-increasing demand for the natural resources needs novel technologies to improve the underground exploration and to produce more crops. The subsurface environment and agricultural lands provide various natural resources such as earth minerals, fossil fuels, metal ores, groundwater, and food. To monitor and improve the production of all of these resources, Internet of Underground Things (IoUGT) is an enabling technology which can provide smart oil and gas fields, smart agriculture fields, and smart seismic quality control [7]. However, implementation of IoUGT is a challenging task due to the harsh underground propagation environment which requires low power and small size underground sensors, long-range communication technology, efficient networking solutions, and accurate localization techniques.

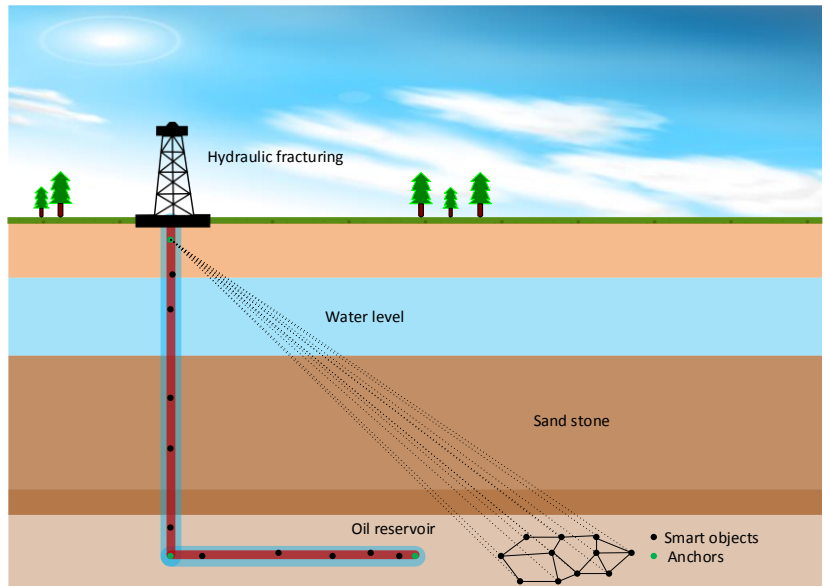

Figure 3 MI-based IoUGT for oil and gas monitoring.

Technological Challenges:

IoUGT consists of a network of buried smart objects consisting of sensors and communication modules. The smart objects can sense and communicate the data to the on-ground base stations. Unlike terrestrial IoT, IoUGT suffers from the harsh underground environment where the communication range is limited. In the following, we discuss the major challenges for IoUGT.

1) Communication Challenges:

Although the work on the development of IoUGT is in the academic research phase, IoUGT can be implemented with different existing communication technologies. For example, for agriculture applications where the sensors are buried at a low depth,

Electromagnetic waves (EM) are better suited [8]. However, in applications such as monitoring of oil and gas reservoirs, magnetic induction (MI) can be a better option due to their low path loss [9]. Figure 3 shows an architecture of MI-based IoUGT for oil and gas reservoirs with smart objects in the reservoir and the anchors attached to the fracturing well. In other applications such as seismic exploration and monitoring underground drilling, acoustic waves are preferred [10]. Table 2 compares the features of potential wireless communication technologies for IoUGT.

| Features | Acoustic waves | RF waves | MI |
| --- | --- | --- | --- |
| Data rate | In tens of bps | In tens of bps | In Kbps |
| Attenuation | Medium | High | Low |
| Cost | High | Medium | Low |
| Transmission distance | In hundreds of meters | Few meters | In tens of meters |
| Interference | Medium | High | Low |

Table 2 Comparison of various wireless communication technologies for IoUGT.

2) Networking and Localization Challenges:

Since the transmission range of each communication technology differs underground but all of them suffer from high transmission loss due to the harsh underground propagation environment and therefore require efficient multi-hop networking solutions. Moreover, most of the IoUGT applications such as monitoring of oil rig, optimized fracturing, and collecting of geo-tagged sensing data, require location information of the deployed smart objects. Few efforts have been made in the past to find the location of buried underground objects. For example, the impact of rocks and minerals on localization accuracy was investigated in [11] for MI-based IoUGT which have shown that at very low frequencies the skin effect is negligible for most of the underground materials. Furthermore, in one of our recent works in [12] we have provided the bound for maximum achievable accuracy for the MI-based IoUGT which takes into account the various network and channel parameters.

Internet of Space Things (IoST):

According to recent news in [13], NASA is planning to start the first human colony on Mars by 2025. To provide global connectivity beyond earth to the humans on

Mars, Internet of Space Things (IoST) is an enabling technology. This way IoST will provide intergalactic communications. Moreover, applications such as virtual reality can be used at earth to walk over MARs by using the IoST. Hence, there has been a growing interest to develop the IoST by the space industry which is mostly in the development phase. Moreover, IoST also provides extended coverage and improve the sensing capabilities of existing satellite networks. For example, Iridium communications have placed 66 satellites (SensorPODs) to provide connectivity to the end users as well as sensing of the outer space and earth [14]. Similarly, Kepler communications have recently launched KIPP satellites (Nano-satellites) in Ku-band for space missions [15]. Currently, the main goal of IoST is to enable global connectivity beyond planet earth and provide sensing capabilities at a low cost. Nanosatellites are envisioned to play a significant role in the development of IoST where inter-satellite communication is also enabled for in-space backhauling and data forwarding as shown in Figure 4.

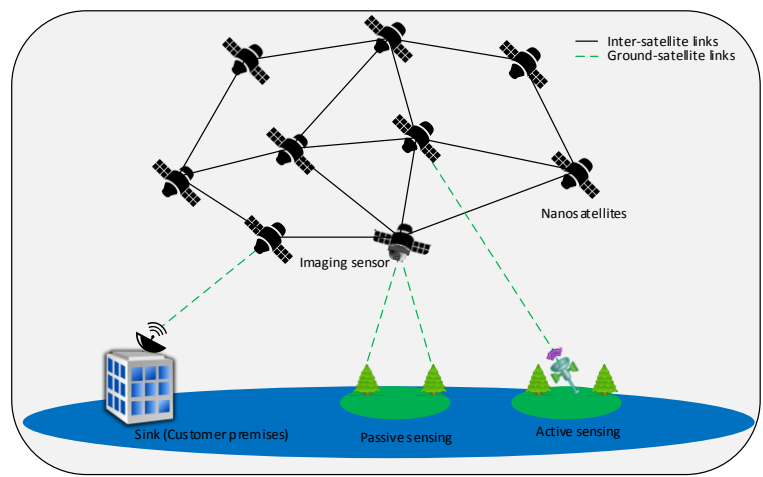

Figure 4 Architecture of IoST.

Figure 5 shows the forecast of nanosatellite launches by 2023 where the number of satellites launches are highly increasing [16]. Moreover, the cyber-physical systems on the ground can be leveraged into the satellite networks to provide a fully integrated system. For example, satellite imaging when combined with sensing information from local in-vehicle sensors can provide optimal navigation path for autonomous driving [17]. Although IoST is going to provide various applications, it is still in the development phase and faces various challenges.

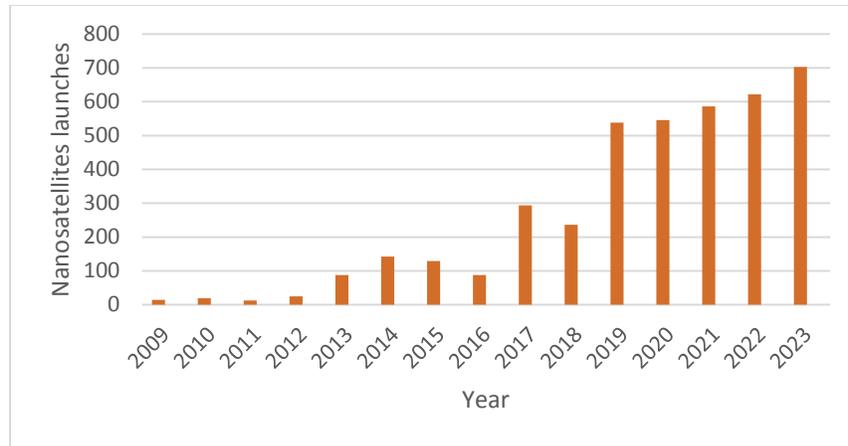

Figure 5 Forecast of nanosatellite launches [16].

Technological Challenges:

The major challenges faced by IoST include resource allocation techniques, long delays, variable network topology, and synchronization [18]. Resource allocation techniques for IoST can jointly optimize the physical-link layer parameters such as operating frequency band, transmission power, bandwidth, number of antennas, and scheduling. As the footprints of nanosatellites are small, it requires the deployment of a satellite constellation for continuous global coverage. Furthermore, the dynamic nature of IoST network requires to develop novel routing protocols for data forwarding. For example, in [18] a software-defined based approach is used for IoST. The establishment of inter-satellite links is also challenging due to the dynamic topology of IoST. Radio frequencies and optical waves are used to establish the inter-satellite links where each technology is having its pros and cons as shown in Table 3.

| Features | RF waves | Optical waves |
|---|---|---|
| Data rate | In hundreds of Mbps | In Gbps |
| Antenna size | Large | Small |
| Cost | High | Low |
| Beamwidth | Wide | Narrow |
| Interference | High | Low |

Table 3 Comparison of wireless communication technologies for inter-satellite links in IoST.

Another major challenge for the IoST is the synchronization among the satellites which can cause a large error for location-based services.

## Conclusions:

We envision that X-IoT is going to play a fundamental role in the future by exploring and connecting the underwater, underground, and space environment. The X-IoT framework can also be extended to fuse other interesting IoT technologies such as the Internet of battlefield things (IoBT) and the Internet of music things (IoMT). However, all these different types of IoT frameworks merging into the X-IoT framework faces various challenges such as communication, networking, localization and security, etc. Moreover, the joint framework of X-IoT will lead to collect an enormous amount of data from multiple types of IoT networks. Therefore, X-IoT will require big data analytics, machine learning and AI to automate the decisions in these IoT frameworks. Also, we believe that the recent advances in the development of smart objects, networking protocols, data analytics tools, and machine learning will shape the future of X-IoT. In short, this exciting area of research opens many important future directions for each category of X-IoT.